\documentclass[preprint, 1p, onecolumn, number, sort&compress, pdftex]{elsarticle}
\usepackage[T1]{fontenc}
\usepackage[latin1]{inputenc}
\usepackage[english]{babel}
\usepackage{amssymb, amsmath}
\usepackage{bm}
\usepackage{textcomp}
\usepackage{color}
\usepackage{lmodern}
\usepackage[bookmarksnumbered=true, bookmarksopen=false, colorlinks]{hyperref}

\newcommand{\eq}{\begin{equation}}                                                                         
\newcommand{\eqe}{\end{equation}}

\journal{Communication in Nonlinear Science and Numerical Simulation}  
\begin{document}
\title{Analytic solutions for irregular diffusion equations with concentration dependent diffusion coefficients} 

\begin{frontmatter}
\author[wigner]{I.F. Barna\corref{imre}}

\cortext[imre]{Corresponding author}
\ead{barna.imre@wigner.hun-ren.hu}
\author[sapientia]{L. M\'aty\'as}

\address[wigner]{Hungarian Research Network, Wigner Research Centre for Physics, Konkoly--Thege Mikl\'os \'ut 29-33, H-1121 Budapest, Hungary}

\address[sapientia]{Sapientia University, Department of Bioengineering, Libert\u{a}tii sq. 1, 530104 Miercurea Ciuc, Romania}

\begin{abstract} 
  We investigate diffusion equations which have concentration dependent diffusion coefficients with physically two relevant Ans\"atze, the self-similar and the traveling wave Ansatz. We found that for power-law 
concentration dependence some of the results can be expressed with a general analytic implicit formulas for both trial functions. 
 For the self-similar case some of the solutions can be given with a formula containing the hypergeometric function. For the traveling wave case
different analytic formulas  are given for different exponents. 
 For some physically reasonable parameter sets the direct solutions are given and analyzed in details.   

\end{abstract}

\end{frontmatter}


\section{Introduction}
The simplest mass or energy transfer is diffusion or heat conduction phenomena which 
is one of the most important mass transport phenomena in nature therefore 
it has an enormous interest in science and engineering as well. The corresponding literature 
has grown enormously over the past 200 years. Without completeness we just mention some 
recent textbooks \cite{gez, dif2,dif3,heat}. 

The simplest diffusion process is the regular one, which is formulated with 
 well-known parabolic partial differential equation (PDE) in the form of 
\eq
\frac{\partial C(x,t)}{\partial t} = D \cdot \Delta C(x,t), 
\label{eq-diff}
\eqe
where $C(x,t) $ is the concentration an D is the diffusion coefficient which is a positive 
real constant and  $\Delta$ represents  the Laplace differential operator in arbitrary dimensions in arbitrary coordinate system. 
Certain boundary conditions belong to equation (\ref{eq-diff}). 
There are definite solutions for finite systems, which often are related 
to engineering applications \cite{Thamby2011,Bennett2013}. 

There are numerous works done in the last decades to find analytic solutions beyond the well-known Gaussian and error functions. 
The best known is the work of  Bluman and Cole \cite{blum} from 1969 who gave an analysis based on a general symmetry 
analysis method giving numerous analytic solutions, some of them are expressible with Gaussian or error functions. As we see they did not present any kind of solutions which looks similar to ours (and which we  present in the following). 

In case of infinite horizon the fundamental solution is the Gaussian 
function which has application in different areas of science. 
In the last year we found additional, physically relevant 
solutions with the help of the self-similar Ansatz 
\cite{laci,imre1} which are a logical generalization of the fundamental solution. These solutions are the multiplication of the 
Gaussian function and the Kummer's M and Kumer's U functions which have an additional free parameter, due to this 
parameter even decaying and oscillatory solutions can be given.  The applied self-similar Ansatz will 
be defined later on in this study in all details. 
The diffusion coefficient in certain cases may be considered constant, 
however there are cases where it may vary \cite{Reif}. 
For two dimensional models important results have been obtained in 
ref.\cite{Claus2001} and for a diffusive-reactive case in ref.\cite{Matyas2005}. Environmental aspects of diffusive very fine particles is discussed in \cite{SaFa2015}.

Diffusion on surfaces is also a significant topic, with possible irregular features \cite{Matyas2004}. 
The chaotic properties of this latter system have been studied in 
\cite{Matyas2011}. 
 
The general form of diffusion equation comes from a conservation law and reads,   
\begin{equation} 
\frac{\partial C}{\partial t} =  \frac{\partial}{\partial x} 
\left( D \frac{\partial C}{\partial x}  \right).   
\label{eq-diff2}
\end{equation} 
In case the diffusion coefficient $D$ depends on parameter $C$, then in general it will also depend on $x$.  
So the case  $D(C[x,t])$ is possible.  
The diffusion coefficient may depend on certain physical quantities, 
and it may vary depending in which phase the system is:       
gaseous, fluid or solid phase. 
In this sense there is a difference between the dependence of 
mass diffusion coefficient and heat diffusion coefficient dependence on 
the parameter $C$. The $C$ stands for the density or concentration in 
case of mass diffusion and $C$ corresponds to temperature in case of 
heat diffusion. 

The investigation of the regular diffusion is just the first step to understand diffusion process in general, however there are much more complex and difficult diffusion processes in nature. 
The diffusion coefficient can have additional dependencies like, time, space of even a tricky combination of both. 
(We examined the case when the diffusion coefficient depend on the function $x/t^{\frac{1}{2}}$)  
In the last years we systematically investigated of such equations and gave new self-similar solutions together with detailed numerical investigations as well. We applied explicit, semi-explicit and implicit numerical schemes to 
solve numerically the corresponding PDEs and compared the evaluated solutions to the exact mathematical ones 
\cite{endre1,endre2,endre3}. We found that in most cases the leapfrog-hopscotch method is the most expedient method to solve PDEs. These kind of diffusion equations with time, space and "time-and-space" dependent diffusion coefficients do have analytic solutions 
with the similar structure the fundamental Gaussian solution multiplied by the Kummer's M and Kummer's U functions or with the Whittaker type functions. 
Due to the extra free parameter different type of decaying solutions 
are always exist with different asymptotics. Some solutions have additional oscillations as well.    

On this way we may go a bit further and ask the question what are the properties of the diffusion 
process when the diffusion coefficient directly depend on the concentration, in the form of   $D(C[x,t])$. These cases cover certain real non-linear diffusion processes described with highly non-linear PDEs.  
In the following we will investigate this question in details. 

\section{Theory and Results}

It is evident that diffusion is in general a three dimensional process beyond Cartesian symmetry, however 
we limit our analysis to a single Cartesian space dependent equation.  
In case we consider the equation of heat transfer, then 
we have for the heat flux $q = -\kappa T_x$, where $\kappa$ is 
the thermal conductivity. If $\kappa$ depends on temperature 
we have for the energy balance equation,        
\eq
\rho c_p   \frac{\partial T}{\partial t} =  \frac{\partial}{\partial x} \left(\kappa(T) \cdot  \frac{\partial T}{\partial x}\right). 
\label{egyen}
\eqe
  
 The function of  $\kappa(C[x,t])$ in principle can be any kind of continuous function with existing first derivative with respect to the concentration $C(x,t)$.  
 
There are numerous studies available in which various the non-linear diffusion (or heat conduction) equations were solved with different methods and means and sometimes analytic solutions are given. 
Without completeness we cite the most relevant studies from the last seven decades.

We should start our reference with the work of Fujita \cite{fujita} from 1952 
who gave analytical solutions when the diffusion coefficient has the form of $D(C) = D(0)/(1-\lambda C)$.  

Later Pattle \cite{pattle} in 1959 
gave solutions with compact support for diffusion from an instantaneous point source in one, two, or three dimensions. 
Philip  \cite{philip} derived exact solutions for the non-linear diffusion equation 
when the concentration has the forms of $D(C) = D_0/(1-\lambda C), D(C) =  D_0/(1-\lambda C)^2 $ and $D(C) = D_0/(1 +2a C +bC^2)$. 

 Boyer \cite{boyer} used a special kind of general self-similar Ansatz with the form of $T(x,t)= U(t)Y(r/R(t))$ and solved the non-linear heat conduction equation when the thermal conductivity was given in the form of $\kappa(T) = (\alpha   + \beta T + \gamma T^2)^{-1} $.   

In 1964 Bankoff \cite{bankoff} investigated heat conduction or diffusion with change of phase and presented numerous methods how the solutions can be derived with series expansion or integral methods. 

Knight and Philip \cite{knight} gave solution for the case of
$D(C) = a(b-C)^{-2}$ with the help of the linearisation of the 
equation.

Tuck \cite{tuck} solved the diffusion equation for  $D(C) = kC^n$  with the constant source boundary condition with the self-similar solution and presented results with compact supports. 

Munier {\it{et al.}} \cite{mun} presented that the  self-similar and the 
partially invariant solutions are identical and introduced the theory of homology with new type of solutions. These are related through the B\"acklund transformation. They acknowledge in the paper, that for $n=-1$ is an exceptional case, where they 
met singularities. In this paper we discuss the case $n=-1$, and we give an explicit continuous solution.  
  
King  \cite{king} solved a cylindrical symmetric  non-linear diffusion equation with the self-similar Ansatz (which is very similar to our) and 
found solutions which can be expressed with the  Airy functions. 

Sadighi and Ganji \cite{said} applied the variational
iteration method and presented analytic results in form of final polynomials. 
 
 Hayek \cite{hayek} presented an exact solution for a nonlinear diffusion equation in a radially symmetric  n-dimensional case in inhomogeneous medium  with the help of the self-similar Ansatz  (Eq.\ref{ans1}) which we also apply. 
 
 Kosov and  Semenov \cite{kos} derived new radially symmetric exact solutions of the multidimensional nonlinear diffusion equation, which can be expressed in terms of elementary functions, Jacobi elliptic functions, Bessel functions, the exponential integral and the Lambert-W function.
 
Luckily to us, it must be strongly emphasized that 
we cannot identify solutions in the above cited 
literature which is exactly the same as our  Eq. (\ref{a_tuti_megoldas}).  
Our personal problem with some of the mentioned studies that after doing the  symmetry 
reduction the mathematical properties of derived ODEs are not investigated and the parameter dependence of the solutions 
are seldom discussed in details. We think that is the crucial point of the whole 
investigation and when that is properly done then much larger part of the scientific community (including applied physicists, chemist or various engineers) can directly profit from these solutions. We think that clear-cut well-explained analytic formulas still have 
high relevance in the understanding of physics.

As an outlook we mention some analytic studies of some non-linear  reaction diffusion equations which are non-linear diffusion equations with one or more extra source terms. 

An exhaustive group classification of such equation were evaluated 
by Dorodnitryn  \cite{dorod} in 1982. 

The non-classical symmetry reduction was done by Arrigo 
{\it{et al.}} \cite{arrigo} in 1994. 

Vijayakumar \cite{vij} presented a study in which he investigated the generalized diffusion equations (the Fisher, Newell-Whitehead and Fitzhugh-Nagumo equations) via the isovector approach and showed analytic results as well. 

Cherniha and Serov \cite{cer}  investigated the more general nonlinear diffusion equations with convection term with the Lie and non-Lie symmetry methods and presented 
additional solutions. However non clear-cut formulas are given with parameter studies.

Reductions and symmetries properties  for a generalized Fisher equation with a diffusion term dependent on density and space 
was investigated by Chulian {\it{et al.}} \cite{chiu}. 

Liu \cite{liu} gave a generalized symmetry classification, and gave the integrable properties with exact solutions for some nonlinear reaction-diffusion equations.   

Qu {\it{et al.}} \cite{qu} applied the conditional Lie - B\"acklund symmetries 
with differential constraints and presented
explicit solutions for a class of  nonlinear reaction-diffusion equations.

It is important to mention here that this is just the simplest (the phenomenological) way to introduce non-linearity into the heat conduction equation. 
We still apply the Fourier law, where the heat flux is equal to the 
temperature gradient times thermal conductivity $q = -\kappa T_x$ 
which has now got an extra temperature dependence $\kappa(T)$. 
There is a mathematically more rigorous method to derive non-linear heat conduction equations which go beyond the Fourier law. 
The heat flux should be approximated with the higher temporal derivatives 
of the temperature. If the second term is considered we arrive to the 
Cattaneo-Vernotte equation \cite{cat1,cat2,ver}, more on 
such kind of heat conduction equations can be found 
in the classical work of  Gurtin and Pipkin  \cite{gur}
or in the  work of Joseph and Preziosi \cite{josprec}. 
Such heat conduction equations have finite signal propagation velocity 
properties. An Euler-Poisson-Darboux type of non-autonomous time-dependent heat conduction equation was derived by Barna and Kersner \cite{imrerobi} which had solutions with a compact support. 
To describe heat pulse experiments "beyond the Cattaneo-Vernotte" models were applied bye Kov\'acs and V\'an \cite{kovvan}.  

Numerical investigation of Eq. (\ref{egyen}) was done by \cite{filip} applying the implicit Euler method. 
In the following we will investigate this equation with the reduction mechanism applying two physically relevant 
trial functions, the self-similar and the traveling wave Ans\"atze.  
It is worth to mention here that this equation is a bit similar to the 
porous media equation which has the form of 
$ U_t = \Delta (U^m)$ where $m > 1$ and was heavily  investigated  in former times \cite{por1,por2,por3}. We have to emphasise that 
if eq. (\ref{eq-diff2}) has an extra source than we arrive to further scientific fields as are  
the Fisher equation \cite{Fisher,Al-Khaled2001}, Turing model \cite{Turing1952} or Swift - Hohenberg equation \cite{Boccaletti1999}  
for nonlinear optics. The diffusion equation has a wide range of 
applicability in science, which also includes the theory of pricing 
\cite{Mazzoni2018,Lazar2014}.  

After dividing by $\rho c_p$ equation (\ref{egyen}) can be also written 
as a diffusion equation,   
\begin{equation} 
\frac{\partial T}{\partial t} =  \frac{\partial}{\partial x} \left(D_h(T) \cdot \frac{\partial T}{\partial x} \right),
\label{egyen-2}
\end{equation}  
where $D_h = \kappa/(\rho c_p)$ is the heat diffusion coefficient. 

Because the heat and mass diffusion equation has the same PDE form, we 
simply consider the general dependence $D = \zeta(C)$  
for (\ref{eq-diff2}), which corresponds to $D_h = \zeta(T)$ 
for (\ref{egyen-2}) and we study the following equation:  
\begin{equation} 
\partial_t C = \zeta(C)_C \cdot C_x^2 + \zeta(C) \cdot C_{xx},  
\label{egyen=3}
\end{equation}  
where the subscripts mean the partial derivations, respectively. 
In the following we study the consequences of such dependence, when the diffusion may vary in terms of the parameter $C$. 

\subsection{Self-similar Analysis}
Let's start with the self-similar Ansatz \cite{por3,Ames1965,sedov} of the form of 
\eq
C(x,t) = t^{-\alpha}f\left( \frac{x}{t^{\beta}}\right) = t^{-\alpha}f(\eta),
\label{ans1}
\eqe
where $f(\eta)$ is the shape function with the reduced variable $\eta$, the two 
self-similar exponents 
$\alpha$ and $\beta$ are responsible for the decay and spreading of the solutions if both have non-negative values. 
In the last decade we generalized this kind of Ansatz 
to multiple spatial dimension and applied it to the Rayleigh-B\'enard
 convection problems \cite{ben1,ben2} or to the heated boundary layer equations \cite{bound}. 
 
 To make and in-depth analysis the functional form of the concentration dependent diffusion coefficient has to be defined. 
We start with the most evident case, the power law dependence: 
\eq
\zeta (C[x,t]) = a \cdot C(x,t)^n \hspace*{0.5cm}  \textrm{where  n}  \varepsilon \mathbb{R} 
\eqe
and the constant $a$ has to role to fix the dimension.
(For simplicity we fix its numerical value to unity.)

If we insert this equation into (\ref{egyen=3})  we get the following equation  
\begin{equation} 
\rho c_p \left(-\alpha t^{-\alpha-1} f - \beta\, t^{-\alpha-1} f'\right) 
=  n t^{-\alpha n-\alpha-2\beta} f'^2 + t^{-\alpha n - \alpha -2\beta} f'',
\end{equation}
where prime means derivation in respect to $\eta$. 

After simplification with $t^{-\alpha}$ one arrives to:  
\begin{equation} 
\rho c_p \left(-\alpha t^{-1} f - \beta\, t^{-1} f'\right) 
=  n t^{-\alpha n-2\beta} f'^2 + t^{-\alpha n -2\beta} f''. 
\end{equation}
Both sides of the equation has the same decay in time if 
\begin{equation} 
-1 = - n \alpha - 2 \beta,
\end{equation}
or $ \alpha = (1- 2\beta)/n $. 

After simplification with $t^{-1}$, and inserting the value of $\alpha$ in the equation, we arrive to the ordinary differential equation (ODE) of 
\eq
\rho c_p \left(-  \left[\frac{1-2\beta}{n}\right]f(\eta) - \beta \eta f(\eta)'       \right)  = n f^{n-1}f'^2 + f^nf''.
\label{ode1}
\eqe
Now we have to make case studies for different $n$s.  

\subsection{Case $n=0$}

If $n=0$ we have the differential equation which corresponds to 
the usual diffusion equation, where the diffusion coefficient is constant.
Although this is considered a relatively known case, here we present 
two very recent results for infinite horizon \cite{MaBa2023a}. 
Beyond the usual Gaussian solution 
\begin{equation} 
C(x,t) = \frac{1}{\sqrt{t}} e^{-\frac{x^2}{4Dt}}, 
\end{equation}
there are further relative simple solutions. 
There is a countable set of even solution relative to the spatial coordinate, the most simple one is (beyond Gaussian) 
\begin{equation}
C(x,t) = \frac{1}{t^{\frac{3}{2}}}  e^{-\frac{x^2}{4Dt}} 
\left( 1 - \frac{1}{2D} \frac{x^2}{t} \right) .
\end{equation}
There is also a countable set of odd solutions relative to the spatial coordinate, a simple one is 
\begin{equation}
 C(x,t) = \frac{x}{t^{\frac{5}{2}}}  e^{-\frac{x^2}{4Dt}} 
\left( 1 - \frac{1}{6D} \frac{x^2}{t}  \right) .
\end{equation}

\subsection{Case $n\neq 0$}

If $n\neq 0$ the equation (\ref{ode1}) is more complex. 
Unlike the regular diffusion equation, we have now three parameters
$\alpha,\beta$ and $n$ and two of them remains free,   
$\alpha = (1-2\beta)/n$. The solution has now the direct form of 
$C(x,t) = t^{\frac{1-2\beta}{n}}f(x/t^{\beta}) $ 
The obtained ODE (\ref{ode1}) has no general closed form solution for arbitrary $\beta$ and n. We will see that only for some special fixed 
$\beta$ and n combinations give us analytic results.  

In the following case $n=-1$ will be considered with special attention, 
because corresponds to a characteristic dependence for dilute systems 
\cite{Reif,Matyas2005}.

\subsubsection{ Case $\beta = 0$}
We may start with the case of  $\beta = 0$ which give us an implicit solution of  
\eq
\int^{f(\eta)} \pm \frac{ n a^{2n}(2+n)}
{\sqrt{-n a^{2n}(2 +n)(2 a^{2+n}\rho c_p -c_1) }}da - \eta - c_2 = 0.
\eqe

It turned out after some algebra, that if all four parameters of the integral $n,c_1, \rho$ and $c_p$ are 
arbitrary rational numbers, there is a definite solution which can be expressed 
with the $_2F_1()$ hypergeometric function \cite{NIST} 
\begin{eqnarray}
     \left(   n(n+2)f(\eta)^{1+2n}\sqrt{1- \frac{2\rho c_p f(\eta)^{2+n}}{c1} } \right)  \times \nonumber \\ 
 _2F_1\left( \frac{1}{2}, \frac{1+n}{2+n}; 1+\frac{1+n}{2+n}; \frac{2\rho c_p f(\eta)^{2+n}}{c_1} \right) \times \nonumber \\ 
\left ( (1+n)\sqrt{-n(2+n)f(\eta)^{2n}[2\rho c_p  f(\eta)^{2+n}- c1] }  \right)^{-1} - \nonumber \\ 
       \eta - c_2 = 0.   \nonumber \\     
       \label{a_tuti_megoldas}
 \end{eqnarray}
Considering the more special case of  $c_1 = 0$ the integral can be given in closed form 
for general $\rho c_p$  and $n$ values, so the implicit equation reads as follows:   
 \eq
\pm \frac{\sqrt{2}(2+n)f(\eta)^{1+2n}}{\sqrt{-n f(\eta)^{3n+2}(2+n)\rho c_p}} -\eta -c_2 = 0 \, .
\eqe
As we can see on this result, that in the denominator of the fraction, 
the argument of the square root is positive if for certain constraints. 
One of the possibilities is if $n$ is a negative number with small absolute value. 
We rise this latter equation to the second power and we get 
\begin{equation} 
f(\eta)^n = - \frac{n\rho c_p}{2(2+n)} (\eta+c_2)^2  . 
\end{equation}  

For $n=-1$ both sides of the equation are positive 
\begin{equation}
\frac{1}{f} = \frac{\rho c_p (\eta+2)^2}{2}   . 
\end{equation} 
This gives for the $C(x,t)$, in case $\alpha=(1-2\beta)/n=-1$:  
\begin{equation}
C(x,t) = t f(\eta) = t \cdot \frac{2}{\rho c_p (x+c_2)^2}  .  
\end{equation}  
One may arrive to this result by applying to equation (\ref{eq-diff2}), 
the standard change of variables $C(x,t)=A(t)\cdot B(t)$, for $n=-1$. 
This relation fulfils the equation (\ref{egyen}), 
however it is divergent for large times. 


\subsubsection{Case $\beta=1$} 

For the second case let's take   $\beta = 1$ and $n = -1$ with the solution of 
\eq
f(\eta) = \frac{c_1^2}{-\rho c_p(1+c_1\eta) + c_1^2c_2e^{c_1\eta}},
\label{sol1}
\eqe

It is clear that for the case $\rho c_p(1+c_1\eta) = c_1 c_1^2 e^{c_1\eta}$ the shape function 
becomes singular at one or two points where the linear equation is touches or intersects the exponential 
equation a nice example, when all four parameters have unit values. We exclude such non-physical solutions 
from our analysis. It is also clear that such solutions arise when the $\rho$ and $c_p$ are much larger than 
the initial conditions $c_1$ and $c_2$.

\begin{figure}
\scalebox{1.0}{
\rotatebox{0}{\includegraphics{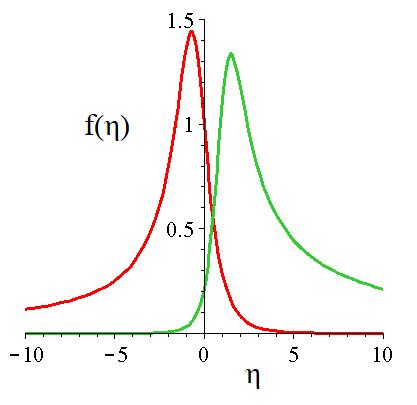}}}
\caption{The solution of Eq. (\ref{sol1}). 
The used parameter sets $\rho\cdot c_p, c_1,c_2$  
for the red and green lines are  
$(1,1,2 )$, and $(1,-2,5)$, respectively.}
\label{egyes}      
\end{figure}

Figure (\ref{egyes}) shows  the solutions of Eq. (\ref{sol1})  for the shape functions for two different parameter sets. 

We analyze the function (\ref{sol1}) by evaluating the derivative of it  
\begin{equation} 
\frac{\partial f}{\partial \eta} 
= \frac{c_1^2 (\rho c_p c_1 - c_1^3 c_2 e^{c_1 \eta})}{(-\rho c_p (1+c_1 \eta)
+c_1^2 c_2 e^{c_1 \eta})^2 }.  
\end{equation} 
As one can see, this function has an extreme value if
\begin{equation} 
\rho c_p c_1 = c_1^3 c_2 e^{c_1 \eta}. 
\end{equation}
This means that this extrema will occur at the value of $\eta$
\begin{equation} 
\eta_* = \frac{1}{c_1} \ln \frac{\rho c_p}{c_1^2 c_2}.
\end{equation}
One can see, that if  
\begin{eqnarray} 
\rho c_p &>& c_1^2 c_2, \,\,\, \eta_* > 0,  \\ 
\rho c_p &<& c_1^2 c_2, \,\,\, \eta_* < 0. 
\end{eqnarray}

\begin{figure}
\scalebox{0.55}{
\rotatebox{0}{\includegraphics{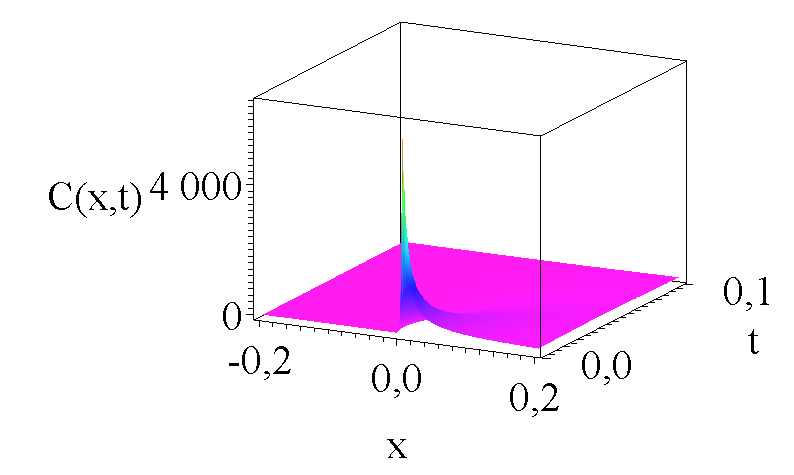}}}
\caption{The solution of Eq. (\ref{C_sol1}), the presented $C(x,t)$ 
function is for $\alpha = 1,  \beta = 1,  \rho = 1, 
c_p = 0.1, c_1 = 15, c_2 = 0.7, $ parameter set,
  respectively.}
\label{kettes}      
\end{figure}

Figure (\ref{kettes}) however presents the $C(x,t)$ total solution in the form of:
\eq
 C(x,t) = \frac{1}{t} \left( \frac{c_1^2}{-c_p \rho (1 + c_1 \rho (x/t) + c_1^2c_2e^{\frac{c_1 x}{t} } } \right). 
\label{C_sol1}
\eqe

As a third mathematical case we found a solutions for the parameter pair of $\beta = -1/2$ and $n= -2$. Unfortunately, the result is a multi-valued implicit formula with real and complex parts.  We tried to tune the parameters $c_1,c_2$ and $\rho c_p$ but cannot found any solution which 
could be interpreted physically e.g. has some reasonable asymptotic for infinite time and space coordinates. 

There are exotic but existing real materials which have temperature dependent heat conduction coefficients. As a first example we may mention 
magnetically aligned single wall carbon nanotube films \cite{nano} where 
the heat conduction coefficient has linear temperature dependence between 
50 and 250 K. 
Our second example is bulk semiconductor at large temperature gradient. 
The authors approximate the heat flux with a sum of higher spatial derivatives of the temperature 
\eq
q = -\kappa_0(T)T_x -  \kappa_1(T) T_{xxx} - \kappa_2(T)
-\kappa_3(T)(T_x)^3,
\eqe
where the first coefficient is $\kappa_0 = 1/T$ \cite{semi}. 
In our present model we cannot take into account the higher terms.

\subsection{Traveling wave analysis}
The second physically relevant trial function which we use is
 the traveling wave Ansatz in the form of:
\eq
 C(x,t) = g(x+ct) =g(\omega), 
\label{ansatz2} 
\eqe
to avoid further misunderstanding we use a different notation for the shape function which is $g$ and for the reduced variable 
which is $\omega $ now.
  
 To make and in-depth analysis the functional form of the concentration dependent diffusion coefficient has to be defined. 
Let's try the most evident case, the power law dependence first: 
\eq
\kappa(C[x,t]) = a\cdot C(x,t)^n \hspace*{0.5cm}  \textrm{where  n, b} \>  \varepsilon \mathbb{R} 
\backslash 0, 
\eqe
$n$ is a free exponent and $a$ is responsible for the proper physical dimension of the thermal conductivity. 
(The numerical value of $a$ is set to unity again.)
After the usual algebraic steps we arrive to the ODE of 
 
\eq
\rho c_p c g' =  a\left( n g^{n-1}g'^2 + g^n g'' \right). 	
\label{ode2}
\eqe
 With the help of Maple 12 we can derive a general implicit formula which contains an 
integral 
\eq
\int_{g(\omega)} \frac{a Z^n}{c_1 a + Z \rho c_p c}dZ  -\omega -c_2 = 0. 
\eqe
Luckily, for $n = -1,0$ and $1$ exist closed form solutions. 
For $n = 0$ we get back the regular diffusion equation with the 
exponential front solution, which is nonphysical. 
For $n = 1$ the solution is the sum of the Lambert W function \cite{NIST} with the pure argument of $\omega$ plus a function of $\omega$.  All together the solution is  
divergent at large x and t arguments.  
For completeness we mention that in the work of Kosov and Semenov \cite{kos} a completely different solution is presented where an exponential function has an argument proportional to $[x^4/t + \text{Lambert W}( x^4/t)]$. We cannot transform the two solutions into each other with finite algebraic steps, so our solution is different to \cite{kos}.  (One may find more about Lambert W function in \cite{lambert}. ) 
Luckily, for $n = -1$  the solutions become simpler and 
we get 
\eq
g(\omega)  = \frac{e^{-\frac{\omega +c_2}{c_1 a  }}}{ c_1 \rho c_p c e^{-\frac{\omega +c_2}{c_1 a  }}-1}.
\label{trav_shape}
\eqe

\begin{figure}
\scalebox{1.0}{
\rotatebox{0}{\includegraphics{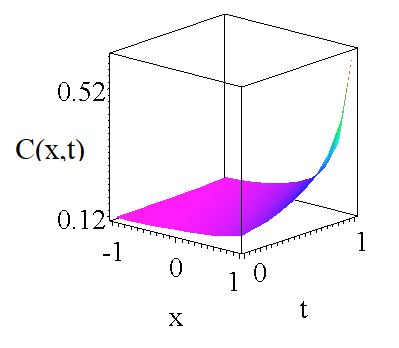}}}
\caption{The concentration function $C(x,t)$ of Eq. (\ref{Cxt_trav}) for the parameter set $c_2=0,\rho \cdot c_p=1,c=1,c_1=10,a=0.1 $.}
\label{negyes}      
\end{figure}

Using the definition of the traveling wave Ansatz the final form 
of the concentration reads:
\eq
C(x,t) = \frac{e^{-\frac{(x+ct) +c_2}{c_1 a  }}}{ c_1 \rho c_p c e^{-
\frac{(x+ct) +c_2}{c_1 a  }}-1}.
\label{Cxt_trav}
\eqe
Figure (\ref{negyes}) shows the $C(x,t)$ concentration function Eq. (\ref{Cxt_trav}) for the set of parameters 
$c_2=0,\rho \cdot c_p=1,c=1,c_1=10,a=0.1 $.


As a second class of functions we may consider a temporal and spatially periodic 
dependence of   
 $ \kappa(T[x,t]) = b \sin(T[x,t])$ 
 unfortunately we cannot derive any solution is a reasonable closed form. Additionally we tried the Lorenzian form of 
$\kappa(T[x,t]) =  \frac{a}{1+T(x,t)^2}$
and the exponential form of $ \kappa(T[x,t]) =  b\cdot Exp(-T[x,t])$ in vain, there are no analytic closed form available. 

\section{Summary and Outlook}
We investigated the highly non-linear diffusion equation where  the diffusion constant (now it is rather a parameter) directly depends on the concentration. Two type of trial functions were used and different functional form were analyzed. We found physically relevant analytic solutions which have power law decays at infinite 
times.  
 In the future - as a natural generalization - we try to investigate 
reactions diffusion equations which are diffusion equations with 
extra source terms on the right hand side.  

\section{Acknowledgment} 
- One of us (I.F. Barna) was supported by the NKFIH, the Hungarian National Research Development and Innovation Office. \\
- The authors declare no conflict of interest. \\
- Both authors contributed equally to every detail of the study.\\
- There was no extra external founding. \\
 - All the evaluated data are available in the manuscript.  

                                        
\end{document}